# Localization Effects in $Bi_2Sr_2Ca(Cu,Co)_2O_{8+y}$ High Temperature Superconductors

C. Quitmann[+], P. Almeras[*], Jian Ma[+], R.J. Kelley[+], H. Berger[*], G. Margiritondo[*], M. Onellion[+]

[+] Deptartment of Physics, University of Wisconsin, Madison, WI.~53706
[*] Institut de Physique Appliqueé, École Polytechnique~Fédérale, CH-1051 Lausanne, Switzerland

Doping $Bi_2Sr_2Ca_1Cu_2O_{8+y}$ with Co causes a superconductor-insulator transition. We study correlations between changes in the electrical resistivity $\rho_{ab}(T)$ and the electronic bandstructure using identical single crystalline samples. For undoped samples the resistivity is linear in temperature and has a vanishing residual resistivity. In angle resolved photoemission these samples show dispersing band-like states. Co-doping decreases $T_C$ and causes and increase in the residual resistivity. Above a threshold Co-concentration the resistivity is metallic ($d\rho_{ab}/dT > 0$) at room temperature, turns insulating below a characteristic temperature $T_{min}$ and becomes superconducting at even lower temperature. These changes in the resistivity correlate with the disappearance of the dispersing band-like states in angle resolved photoemission. We show that Anderson localization caused by the impurity potential of the doped Co-atoms provides a consistent explanation of all experimental features. Therefore the $T_C$ reduction in 3d-metal doped high-temperature superconductors is not caused by Abrikosov Gor'kov pairbreaking but by spatial localization of the carriers. The observed suppression of $T_C$ indicates that the system is in the homogenous limit of the superconductor-insulator transition. The coexistance of insulating ($d\rho_{ab}/dT < 0$) normal state behavior and superconductivity indicates that the superconducting ground state is formed out of spatially almost localized carriers.

Dr. C. Quitmann
Dept. of Physics
University of Wisconsin
1150 Univ. Ave.
<u>MADISON, WI. 53706</u>
 U S A
phone: + 608-263 6829
fax:     + 608-265 2334
email: **quitmann@macc.wisc.edu**

## 1. Introduction

High-temperature superconductors (HTSC) are well known to be close to a superconductor-insulator transition [1-3] that can be achieved by cation doping. However, the nature of this transition is not well understood. Particularly puzzling are the results of 3d-transition metal doping. All 3d-transition metal dopants lead to a superconductor-insulator transition., but doping with non-magnetic Zn is more effective than doping with magnetic Fe, Co or Ni [3].

Superconductor-insulator transitions have long been studied in low-temperature superconductors [4]. Here two cases can be distinguished: homogenous ultra-thin films (Pb) [5] and granular cluster systems (Al:Ge) [6]. In both cases, one starts from a superconducting ground state with a macroscopic wavefunction $\Psi(r) = \Psi_0 \cdot \exp(i \cdot \varphi)$. When approaching the superconductor-insulator transition by varying either the film thickness or the metal to non-metal fraction, this macroscopic wavefunction is destroyed. In homogenous systems, this happens through scattering processes which reduce the amplitude $\Psi_0$ . These scattering processes lead to a continuouse decrease of the transition temperature $T_C$ and the superconducting gap $\Delta$ with increasing disorder. In the granular systems the phase coherence between superconducting clusters is broken. In these granular systems the $T_C$ is given by the $T_C$ of the metallic clusters, which is independend of the metal to non-metal fraction. Thus the onset of the transition is independent of composition. The transition width, however, increases as the superconductor-insulator transition is approached due to the˜system.

Our study has two main results: We show that in Co-doped $Bi_2Sr_2Ca_1Cu_2O_{8+y}$ (Bi-2212) superconductivity and localization coexist; we show that the cause of the localization of the single particle states is Anderson localization. The idea that the superconductor-insulator transition is caused by disorder has previously been suggested by different authors [7-9]. Our report is new in two respects. It is the first report linking the changes observed in the resistivity to changes in the electronic dispersion, and we establish for the first time the coexistence of localization and superconductivity in a homogenous system such as the HTSC. Through the combination of angle resolved photoemission and resistivity measurements we establish (below) the presence of the three characteristic features of Anderson localization, which are the decrease of $T_C$, the transition towards insulating behavior ($d\rho_{ab}/dT < 0$) and the disappearance of delocalized, band-like states. Coexistence of localization and superconductivity has previously only been observed in granular systems [6,10] for which there are also numerous˜theoretical studies [11-14].

This investigation was conducted on the HTSC $Bi_2Sr_2Ca_1Cu_2O_{8+y}$ doped with Co. This system is known to exhibit a superconductor-insulator transition [3], and there is theoretical evidence that disorder effects are much stronger in HTSC than in conventional metals [15,16]. Co acts as an impurity because it has a different electronic structure. This will affect the hopping matrix element between Co and the neighboring O-atoms, and thus disturb the periodicity of the $CuO_2$-plane.

HTSC are well suited for this investigation because, due to their layered structure, they are quasi-two-dimensional. Experimentally one observes a semiconducting resistivity ($d\rho_c/dT < 0$) for the resistivity along the c-axis while the resistivity is metallic in the ab-plane ($d\rho_{ab}/dT > 0$). The resistivity anisotropy is about $\rho_c / \rho_{ab} \cong 10^4$ for the carrier concentration used here [17].

The impurity concentrations used in this study are below 2at%. Therefore the separation between impurities (~3nm at 2at%) is much larger then their diameter (~0.1nm). We are thus well below the percolation threshold in 2-dimensions and can expect to be in the homogenous limit where the amplitude of the wavefunction $\Psi_0$ becomes reduced through the scattering processes.

## 2. Experimental

The samples were grown by the conventional self flux method. Their composition was determined with electron probe micro analysis (EPMA) (Cameca CAMEBAX SX-50). Their size is typically $2 \times 2 \times 0.1$ mm$^3$. For single crystals our solubility limit for Co is approximately 2at%. Above 2at% we observe second phases in the˜%%EPMA.

We report on measurements of three samples, one undoped reference sample and two Co-doped samples. The Co-doped samples have Co-contents, as determined by EPMA, of 1.57at% and 1.60at%. Given the uncertainty of the EPMA measurement of 0.3at% they have almost the same Co-content. However their resistivity and electronic structure are very different. In the following they will be called sample **1** and sample **2** respectively.

Angle resolved photoemission (ARUPS) experiments were performed at the 4m normal incidence monochromator of the Synchrotron Radiation Center in Stoughton, Wisconsin with a photon energy of 21eV. After cleaving the samples in UHV ($1 \times 10^{-10}$torr) low energy electron diffraction (LEED) was performed in-situ to check the surface and to orient the sample. The samples were oriented with the a-axis ($\Gamma$-X) parallel to the photon electric field. For the Bi-2212 system this is the direction of the Bi-O bond which does not show the superlattice modulation [18-20] and it is at $45^0$ to the Cu-O-Cu bond direction. All spectra were taken at room temperature with a combined energy resolution of 120meV. The photon angle of incidence was $45^0$. All angles are measured with respect to the surface normal. The binding energies of all spectra are referred to the Fermi energy of a Au-film located next to the sample.

Resistivity measurements were performed on the same samples after taking them out of the chamber. Current and voltage leads were attached by silver epoxy onto the a,b-plane of the crystals to measure the a,b-plane resistivity, $\rho_{ab}(T)$. The absolute value of the resistivity has an uncertainty of about 15% due to the difficulty of determining the geometry factor of the samples.

### 2.1. Electrical resistivity

Figure 1 illustrates the electrical resistivity $\rho_{ab}(T)$ of a pure Bi-2212 and the two doped samples **1** and **2**. The pure sample shows the well known linear resistivity $\rho_{ab}(T) = \rho_0 + a*T$ [21] with virtually zero residual resistivity $\rho_0$. The superconducting transition temperature is $T_{C,mid} = 91$ K. Sample **1** still shows a resistivity linear in temperature but with a residual resistivity of $\rho_0 = 43$ $\mu\Omega$cm. This indicates an increase of the scattering rate $\tau$ caused by scattering from the Co-impurities. At 100K the resistivity is a factor of 2 higher than for the pure sample. For sample **2** we observe a qualitative change. The resistivity at room temperature has increased by a˜factor of 6 compared to the pure sample. Even more important, it decreases slightly in going to lower temperatures before reaching a minimum at $T_{min} = 190$ K and then increases toward lower temperatures. At still lower temperatures, we observe a superconducting transition with $T_{C, mid} = 66$K. The negative temperature coefficient of the electrical resistivity between $T_{min} = 190$ K and $T_C$ is a clear indication of insulating behavior caused by localization of the charge carriers. This is the first indication of Anderson localization.

In the temperature range between $T_C$ and $T_{min}$ the charge carriers are transported through thermally activated hopping between localized single particle sites [22]. Such hopping process are frozen out as the temperature decreases because the thermal energy available to hopp to an empty site becomes smaller. Above $T_{min}$ the thermal energy is sufficient to excite the carriers across the mobility gap into delocalized states [9]. Therefore the size of the mobility gap in this sample is roughly 190 K or 16 meV. A very similar value is found when the superconductor-insulator transition is introduced by changing the carrier density [9] rather than by introducing impurities.

It is important to realize that the minimum in the resistivity can not be explained by sample inhomogeneities. An inhomogeneous sample results in a parallel resistor network of an insulating and a metallic (superconducting) material. In such a parallel network, the resistivity is dominated by the component with the lower resistivity. Thus, one would observe a maximum in $\rho(T)$ for the temperature where both materials have comparable resistivities. Because the experiments were done on single crystals, grain boundaries can also be excluded as a possible origin of the minimum in $\rho(T)$.

The Co-doping affects not only the normal state but also the transition temperature $T_C$. It decreases from $T_C = 91$ K in the pure material to $T_C = 76$ K in sample **1** and $T_C = 66$ K in sample **2**. This decrease is a characteristic feature of homogenous systems. If the system was inhomogenous one would expect the granular superconducting regions to have the same $T_C$ as the pure material. Only their coupling would be perturbed because of intermediate nonsuperconducting regions. This would lead to a broadening of the transition but leave the onset virtually unchanged. The Co-doped Bi-2212 is therefore in the homogenous limit of a superconductor-insulator transition. Our conclusion is supported by photodoping experiments [23]. Here the carrier density is increased by illuminating the sample with a laser beam. This changes the carrier density without the chemical disorder inherent to chemical doping. In these experiments the width of the transition depends on the illumination power (i.e. the carrier density) but the onset for the superconducting transition is unchanged. Photodoping thus leads˜to a granular system.

The relative decrease of $T_C$ between the pure Bi-2212 and sample **1** is: $\Delta T_C/\Delta x \sim 9$ K/at%. Between sample **1** and sample **2** the reduction is much stronger, at least: $\Delta T_C/\Delta x > 30$ K/at%. The initial suppression rate is a little higher then the suppression observed in polycrystalline material which is $\sim 6$K/at% for Co and for Zn [3]. The drastic decrease of $T_C$ going from the metallic sample **1** to the insulating sample **2** is what one would expect when one crosses the mobility edge and reaches the region where the single particle excitations are localized [11]. It is the second indication of˜Anderson localization.

The coexistence of superconductivity and localization has so far only been seen in low-$T_C$ granular systems such as Al:Ge [6] and In:O [10]. Our Co-doped HTSC samples are the first to show such coexistence in a homogenous material. This is of particular interest for theoretical models because it shows that even in the situation where the single particle excitations are localized, the attractive interaction persists and is not completely overwhelmed by the now poorly screened Coulomb repulsion. Therefore, to assume the existence of a pairing interaction even in the insulating phase remains a reasonable starting point for theoretical models of the superconductor-insulator˜transition.

## 2.2. Angle resolved photoemission

Figures 2, 3 and 4 illustrate ARUPS spectra for the identical three samples as in figure 1. We show spectra along two main symmetry directions Γ-M (Cu-O-Cu bond direction) and Γ-X (**a**-axis). The spectra are comprised of three contributions: a decaying background (caused by the valence band); a relatively broad (FWHM ~ 200 meV) dispersing state of Gaussian shape; and elastically scattered electrons forming a Fermi-Dirac distribution. In the following we will focus on the dispersing band because it provides information about the delocalized states.

In the undoped sample (figure 2) the dispersing band is strong along the Γ-X direction. The band becomes visible above the background at $\theta = 8^0$, then disperses towards the Fermi energy $E_F$ and crosses it at $\theta = 14^0$. This crossing point of the band is important because it indicates the size and topology of the Fermi surface. In the absence of changes in the topology, which are not expected upon doping at this low doping level, the crossing point is directly related to the carrier density. Along the Γ-M direction we also observe a dispersing state but its intensity is weaker. The band disperses towards $E_F$ and then remains close, but below, $E_F$ for a significant portion of the zone [24]. This extended region with almost no dispersion causes a near singularity in the density of states along the Γ-M direction, similar to earlier reports on $Y_1Ba_2Cu_3O_{7-x}$ [25,26].

The data for sample **1** are illustrated in figure 3. Along Γ-X we also observe a dispersing band-like state, although its intensity is significantly reduced compared to the pure sample in figure 2a. The dispersion is, within the error bars, identical to that of the undoped sample. The band crosses the Fermi surface at $\theta = 14^0$, the same location for which it crosses in the pure sample. This indicates an unchanged carrier concentration. Because Co is a $2^+$ ion such a change would not be expected. Hall effect data also˜indicate that there is no change in the carrier concentration [3].

For the Γ-M direction the spectra in Fig. 3(b) are similar those of the pure sample in figure 2b. The intensity of the dispersing state has decreased only slightly and the band still shows a dispersion towards the Fermi surface, without crossing it. There is less reduction in intensity of the band-like state in the Γ-M direction than in the Γ-X direction. A possible reason is that due to the near singularity along this direction [24] the carrier density is larger, which makes the screening of the Co-impurity potential˜more effective.

The data for sample **2**, which exhibited localized single particle states in the electrical resistivity, are illustrated in figure 4. These spectra are qualitatively different from those of the pure material. Along both symmetry directions we only observe the decaying background of the valence band and a Fermi edge. The dispersing band-like state is not observed. Considering our signal to noise ratio this implies a reduction of the band-like states by more than 90%. However, we still observe a Fermi edge, which is as sharp as in the other samples. There is no indication of a semiconducting gap opening in the normal state of this sample. This excludes a change in the carrier density as the cause for the observed upturn in the resistivity below $T_{min}$ and supports the idea of spatial localization of the carriers. When the single particle states become localized in real space due to disorder, the wavevector **k** is no longer a good quantum number. Therefore one can no longer observe a dispersing band-like state in the sample where the single particle excitations are localized. This is the third characteristic feature of Anderson localization.

### 3. Interpretation

3.1. Anderson localization as cause for MIT

Doping of Co on the 1-2at% level into Bi-2212 leads to significant changes in the normal state electrical resistivity. The absolute resistivity increases and the superconducting transition temperature decreases markedly. Above a threshold value of approximately 1.6at% the samples show insulating $(d\rho_{ab}/dT < 0)$ behavior. The Co-doping also causes the disappearance of the delocalized band-like states seen in ARUPS, but does not affect the elastically scattered states which form a Fermi edge in all samples. We have therefore observed the three characteristics for a superconductor-insulator transition caused by Anderson localization.

The transport properties of HTSC doped with 3d-elements have been studied since the early days of HTSC. Many authors have related the observed decrease of $T_C$ with doping to Abrikosov-Gor'kov (AG) pairbreaking [27] caused by the magnetic moment of the dopant. While this is at first sight an attractive explanation, it leaves a number of questions unanswered. In the case of AG-pairbreaking there is no reason to expect significant changes in the normal state properties. However, investigations have observed a very strong increase in the resistivity and a change to insulating behavior [28]. Assuming AG-pairbreaking, it is also difficult to understand that doping with non magnetic Zn has any effect on $T_C$; Zn is actually more effective in reducing $T_C$ than is Fe, Co or Ni [3].

In this paper we have proposed a completely different mechanism, namely Anderson localization through the impurity potential of the dopand. This mechanism accounts for the spectroscopic data, the transport data and further provides an explanation for the results of Zn-doping.

Anderson localization refers to the spatial localization of single particle states in a non periodic potential [29]. In a non periodic potential Bloch's theorem is no longer valid and therefore the single particle states need not be delocalized. As Anderson showed, in a random, disordered potential, states in the tails of the band become localized. Because of the random potential, they cannot find neighboring sites with energy levels appropriate to match the boundary conditions that would cause a hybridization and delocalize the states. Because this is an effect of quantum coherence, it can occur at very low levels of disorder, where classical scattering would not lead to insulating behavior. In fact, in a strictly 2-dimensional system of finite size all states are localized for infinitesimal disorder [30-33]. In three dimensions, a finite level of disorder is necessary to cause localization of all states [30-33].

HTSC have a layered quasi two dimensional structure. This crystal structure is also reflected in the electronic structure, which is extremely anisotropic [17, 34, 35]. They are thus close to the marginal dimension of d=2 and it is reasonable to expected that in HTSC's even small amounts of impurities can lead to localization. For impurity concentrations below the critical limit the single particle states are still extended. However, in contrast to a perfect Bloch-wave, they have a characteristic decay length, the localization radius $a_H$. This is analogous to the mean free path concept of the Drude model and leads to an residual resistivity $\rho_0$. The critical disorder for the metal-insulator transition depends on the details of the bandstructure and the disorder of the system and has not yet been predicted for real systems.

Other possible explanations for the observed changes in the electrical resistivity and the electronic structure are a Mott-transition [22], or the Kondo-effect [36, 37]. In the case of a pure Mott-transition, the Coulomb repulsion between carriers on the same site leads to an energy splitting between singly and doubly occupied sites. This splitting causes a gap in the electronic spectrum if the carriers do not interact because of their large separation. When they come closer together, due to an increase in density for example, the individual states overlap and form a band, which leads to delocalized states and metallic conduction. For the present system, this possibility can be excluded because the carrier density, and thus the carrier separation, does not change. This is evident from the observed Fermi-surface crossings of the band-like states and from the Hall effect measurements [3]. We also find no evidence of any semiconducting gap in the density of states for the samples that show the insulating upturn in $\rho_{ab}$ .

The Kondo-effect describes a resonance in the scattering crossection of delocalized states by localized magnetic moments [36, 37]. This resonance occurs below a characteristic temperature $T_K$ and leads to a logarithmic increase in the electric resistivity. In the present case the Co-atoms could provide the local moment. If the Kondo effect was responsible for the observed behavior the characteristic temperature would have to be identified with $T_{min}$ , the crossover between insulating and metallic behavior. The ARUPS spectra were taken at room temperature, significantly above $T_{min}$. At room temperature, even sample **2** is metallic and there should be no Kondo effect. Therefore, one would expect to still see a dispersing band at room temperature if the Kondo effect was responsible for the upturn in the resistivity. This is in contrast to the results illustrated˝in figure 4.

## 3.2. Coexistence of Anderson localization and superconductivity

Sample **2** exhibits an increase in the electrical resistivity below $T_{min} = 190$ K, caused by localization of the carriers, and a superconducting transition at $T_{C,mid} = 66$ K. Consequently the superconducting many body wavefunction must be composed of spatially localized single particle states.

Such coexistence of localization and superconductivity is very unusual. Experimentally it has only been observed in granular systems such as Al:Ge [6]˝and In:O [10]. Theoretically it has been studied by a variety of authors [11-14,38] who showed that Anderson localization and superconductivity are not mutually exclusive. Anderson's theorem is still valid in a narrow region on the insulating side of the superconductor-insulator transition [11,12]. In the localized region of the insulator-superconductor transition the density of states can no longer be approximated by a spatial average, but must be considered a local quantity $N(E, \mathbf{r})$. Ma and Lee [11] showed that a superconducting wavefunction can be formed provided there are several localized states within an energy range equal to the superconducting gap $\Delta_0$ of the material:

$$a_H^{\,d} \cdot \Delta_0 \cdot \langle N(E_F, \mathbf{r}) \rangle \quad \gg \quad 1$$

Here d is the dimension, $\Delta_0$ is the superconducting gap, and $<N(E_F, \mathbf{r})>$ denotes the density of localized electronic states averaged over an energy region of the size $\Delta_0$. Taking the localization length on the insulating side of the superconductor-insulator transition from scaling theory they estimated the width of the coexistence˝region $(n_c-n)/n_c$ as

$$(1-n/n_c)^\nu \sim (E_F / \Delta)^{1/d}.$$

Here d is the dimensionality, $n_c$ is the critical concentration for the superconductor-insulator transition and $\nu$ is the critical exponent. Assuming d=2 and taking $\nu = -1$ [9] and values of $E_F \cong 400 meV$ and $\Delta \cong 16 meV$ [39] we obtain a coexistence region of $(1-n/n_c) \sim 20\%$. Using the value of $n_c = 1.6 at\%$ for the critical concentration in the case of Co-doping this yields a coexistence region of only $\sim 0.3 at\%$ Co. The relative width of $(1-n/n_c) \sim 20\%$ is a factor of two larger than that of classical low temperature superconductors where Ma and Lee estimated it to be $\sim 10\%$ [11]. Although both the larger gap $\Delta_0$ and the lower Fermi energy $E_F$ in HTSC favor the coexistence of localization and superconductivity, these effects are partially cancelled by˝the lower dimensionality d=2 as compared to d=3 in the granular systems. The narrow coexistence region is consistant with our observation that sample **1** and sample **2** have very similar Co-content but very dissimilar properties.

## 4. Conclusion

Co-doping in Bi-2212 causes an increase of the normal state resistivity and, above a threshold concentration of about 1.6at%, a superconductor-insulator transition. As is characteristic of homogenous materials, the transition˝temperature $T_C$ is suppressed by the impurity doping. The changes in the transport properties are accompanied by the disappearance of the band-like states in angle resolved photoemission spectra. The data can be interpreted in the framework of Anderson localization of the carriers by the impurity potential of the Co-atoms. The data indicate that the perturbation of the periodic potential induced by the Co-atoms is more significant then the pairbreaking effect of the magnetic moment. It thus provides an understanding why even nonmagnetic impurities, such as Zn, destroy superconductivity in these compounds. For certain compositions the macroscopic superconducting wave function is formed out of carriers that are almost spatially localized in the normal state.

## Acknowledgements

We have benefitted from discussions with Hans Kroha and Patrick A. Lee. This work was supported by the National Science Foundation both directly and through support for the Synchrotron Radiation Center, and by École Polytechique Federale and the Fonds National Suisse de la Recherche Scientific. One of us (C.Q.) is grateful for a grant by the Deutsche Forschungsgemeinschaft.

**Figure Captions**

Figure 1 : Electrical resistivity $\rho_{ab}$(T) within the a-b plane for three Bi-2212 samples. A pure sample with no Co ($T_C$ = 91K), Sample **1** (1.57±.3) at% Co ($T_C$ = 76 K) and Sample **2** (1.60±.3) at% Co ($T_C$ = 66K). The arrow indicates the crossover temperaure between metallic and insulating behavior.

Figure 2 : Angular resolved photoemission spectra for pure Bi-2212 taken at T=300K along a.) Γ-X direction and b.) Γ-M direction. Note the dispersing band-like state in both˝directions.

Figure 3 : Angular resolved photoemission spectra for sample **1**

taken at T=300K along a.) Γ-X direction and b.) Γ-M direction. Note the reduction of the dispersing band-like state, especially˝in the Γ-X direction.

Figure 4 : Angular resolved photoemission spectra for sample **2**

taken at T=300K along a.) Γ-X direction and b.) Γ-M direction. Note the complete absence of a dispersing band-like state.